\documentclass[prb,showpacs,showkeys,lengthcheck]{revtex4-1}
\usepackage{amssymb}
\usepackage{amsmath}
\usepackage{amsfonts}
\usepackage{graphicx}%
\usepackage{epstopdf}
\usepackage{sidecap}
\sidecaptionvpos{figure}{c}
\usepackage[note-name=, use-sort-key = false]{notes2bib}
\begin{document}
\preprint{HEP/123-qed}
\title{Optical Study of BaFe$_2$As$_2$ under pressure: Coexistence of spin-density-wave gap and superconductivity}

\author{E. Uykur$^{*,1}$, T. Kobayashi$^2$, W. Hirata$^2$, S. Miyasaka$^2$, S. Tajima$^2$, and C. A. Kuntscher$^{\dagger,1}$}

\affiliation{$^1$Experimentalphysik 2, Universit\"{a}t Augsburg, D-86159 Augsburg, Germany}

\affiliation{$^2$Department of Physics, Graduate School of Science, Osaka University, Osaka, 560-0043, Japan}


\begin{abstract}

Temperature-dependent reflectivity measurements in the frequency range 85 - 7000 cm$^{-1}$ were performed on BaFe$_2$As$_2$ single crystals under pressure up to $\sim$5 GPa. The corresponding pressure- and temperature-dependent optical conductivity was analyzed with the Drude-Lorentz model to extract the coherent and incoherent contributions. The gradual suppression of the spin-density wave (SDW) state with increasing pressure and the appearance of the superconducting phase coexisting with the SDW phase at 3.6~GPa were observed. At 3.6~GPa, the reflectivity reaches unity below $\sim$95~cm$^{-1}$ indicating the opening of the superconducting gap and shows a full gap tendency at 6 K.

\end{abstract}

\volumeyear{year}
\volumenumber{number}
\issuenumber{number}
\eid{identifier}
\date{\today}
\startpage{1}
\endpage{2}
\maketitle

\section{Introduction}

The discovery of the iron-pnictide superconductors \cite{Kamihara2008} has attracted a lot of attention in the condensed-matter research, and a tremendous amount of work has been performed in a very short time after their discovery. It is now widely accepted that these materials undergo a structural (\textit{T}$_S$)
and magnetic (\textit{T}$_N$) phase transition during cooling down from room temperature, followed by each other, in their parent compound phases. Chemical substitution or external pressure suppresses the magnetically ordered (SDW) state, and a superconducting (SC) region emerges with a critical temperature as high as \textit{T}$_c$ $\approx$ 30 K  \cite{Chen2008, Rotter2008, Zhao2008, Chen2009, Nandi2010, Chu2009}. Either hole doping, electron doping or isovalent doping induces a SC state; however, different SC gap structures were found, demanding detailed studies on the SC state induced by external pressure in the parent compounds. Besides the SC state itself, the interplay between the SDW and the SC state is of particular interest.

Among the iron-pnictides systems, the so-called 122 family ($A$Fe$_2$As$_2$ with $A$: Ba, Sr, Ca, Eu) is the interest of the present study. In terms of chemical doping this class of iron-pnictides is extensively studied; moreover, a fare share of pressure studies was performed compared to other classes. The isovalent doping case (e.g., P-doping of As-sites) is especially interesting, since superconductivity with a $T_c$ as high as for the electron- or hole-doped cases is observed, despite the unchanged carrier concentration. Intuitively, one could simply think of P-doping acting as a chemical pressure, affecting the physical properties in a similar fashion like external pressure. However, studies show that the two most discussed structural parameters in iron-pnictides, namely the pnictogen height and the bond angle, behave oppositely with P-doping and pressure \cite{Kimber2009, Kasahara2010}, demanding further investigation.

The pressure-induced SC phase in the 122 class of pnictides was mostly studied by transport probes \cite{Torikachvili2008, Torikachvili2008a, Alireza2009, Kotegawa2009, Ishikawa2009, Igawa2009, Kurita2011, Colombier2009}: Pressure causes the gradual suppression of the SDW state and the emergence of superconductivity in the parent compounds, whereas for the doped iron pnictides pressure increases $T_c$ \cite{Ahilan2008, Drotziger2010, Klintberg2010}. In contrast, very limited work on the iron pnictides has been done with infrared spectroscopy under pressure \cite{Okamura2013, Baldassarre2012}, focusing mainly on the SDW state. None of these studies was able to demonstrate the existence of the SC state. This is the first pressure-dependent optical study on BaFe$_2$As$_2$ for a temperature range which is suitable for the observation of the superconductivity (previous study was performed down to 110~K only \cite{Baldassarre2012}), with the so-far lowest energy measured under pressure. Our data confirm the gradual suppression of the SDW state under pressure. Moreover, the emergence of a SC state with a full gap characteristic at high pressure is observed for the first time. Our results reveal the coexistence of these two states below the SC transition, with the SDW state possibly being a competing order to the superconductivity. We also compare these findings with those for the isovalent P-doping case.

\section{Experimental}

\begin{figure}[t]%
\centering
\includegraphics[scale=1]{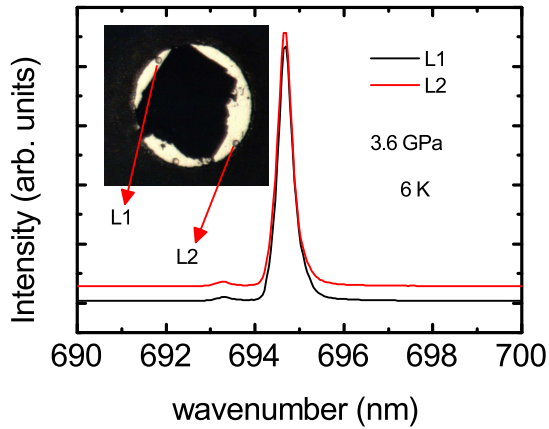}%
\caption{(color online) Ruby luminescence spectra measured at two different location in pressure cell at 3.6 GPa (where, we observe superconductivity). Spectra are given with a $y$-axis offset for clear inspection.}%
\label{hydrostaticity}%
\end{figure}

BaFe$_2$As$_2$ single crystals were grown by the self-flux method and annealed as described elsewhere \cite{Nakajima2010, Nakajima2011}. They were polished and cut into small pieces with a typical size of $\sim$250 x 250 x 60 $\mu$m$^3$. Samples are placed into a type IIa diamond anvil cell (DAC) \cite{Keller1977} and finely ground CsI powder was used as quasi-hydrostatic pressure transmitting medium. The pressure in the DAC was measured {\it in situ} with the ruby luminescence method \cite{Mao1986} by using two different ruby balls at different locations. The luminescence spectra are depicted in Fig.~\ref{hydrostaticity} taken at 3.6 GPa and 6 K (i.e., under conditions where we observe superconductivity) for two different locations as marked in the inset. The ruby R1 line is sharp and the pressure difference for the two locations is below 0.05 GPa.

Reflectivity measurements were performed from $\sim$85 to 7000~cm$^{-1}$  at various temperatures from 6 to 300 K and for pressures up to $\sim$5~GPa with a system described elsewhere \cite{Kuntscher2014}. The reflectivity spectra are measured at the sample-diamond interface (denoted as R$_{s-d}$), using the intensity reflected from the CuBe gasket inside the DAC as reference. Since above 7000 cm$^{-1}$ the reflectivity spectra show no significant pressure or temperature dependence, the ambient-condition spectrum was used for further analysis.

\begin{figure}[t]%
\centering
\includegraphics[scale=1]{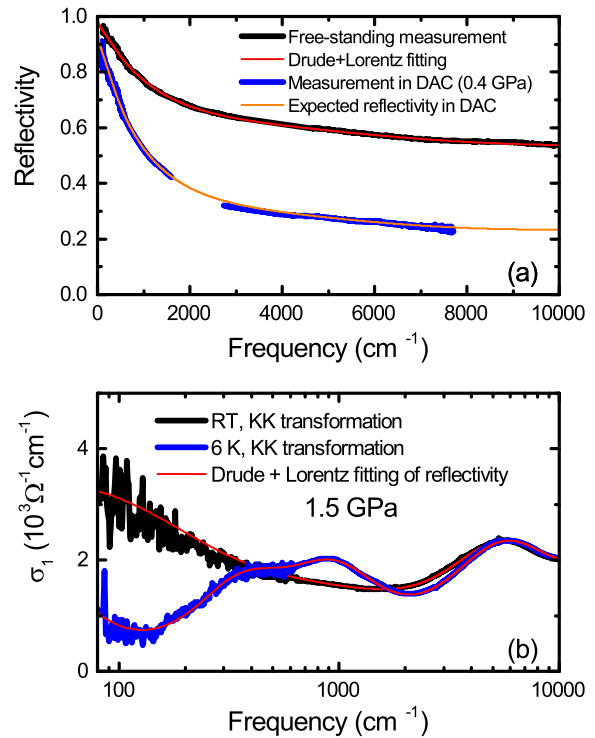}%
\caption{(color online) (a) Free-standing and low-pressure reflectivity spectra at room temperature. Drude+Lorentz fitting of the free-standing reflectivity spectrum and the simulated reflectivity spectrum in the DAC are also plotted. (b) The optical conductivity spectra calculated with the Kramers-Kronig analysis of the measured reflectivity and the Drude-Lorentz fitting of the reflectivity for the comparison.}%
\label{expectedinDAC}%
\end{figure}

For illustrating the difference between reflectivity spectra at the sample-diamond interface compared to sample-vacuum interface, the measured reflectivity spectra for both types of interfaces are plotted in Fig.~\ref{expectedinDAC}(a).
Due to the difference of the refractive index of diamond and vacuum, the measured spectra show a clear difference. Moreover, the expected spectrum in the DAC has been simulated according to the results of the Drude-Lorentz fitting of the free-standing reflectivity measurements. As can be seen from Fig.~\ref{expectedinDAC}(a), the measured reflectivity spectrum in the DAC is in good agreement with the simulated reflectivity spectrum, despite the small pressure offset (0.4 GPa).

The data around 2000 cm$^{-1}$ is affected by the multiphonon absorption in the diamond anvil, and therefore, this energy range has been cut out. This energy region has been extrapolated with the Drude-Lorentz fitting of the reflectivity spectra with the low energy extrapolations. The optical conductivity spectra has been obtained by Drude-Lorentz simulation of the measured reflectivity data, which is in good agreement with the spectra based on a Kramer-Kronig (KK) analysis \cite{Pashkin2006} of the measured reflectivity data. An example to the agreement has been given in Fig.~\ref{expectedinDAC}(b).

\section{Results and Discussion}

\begin{figure}[t]%
\centering
\includegraphics[scale=0.95]{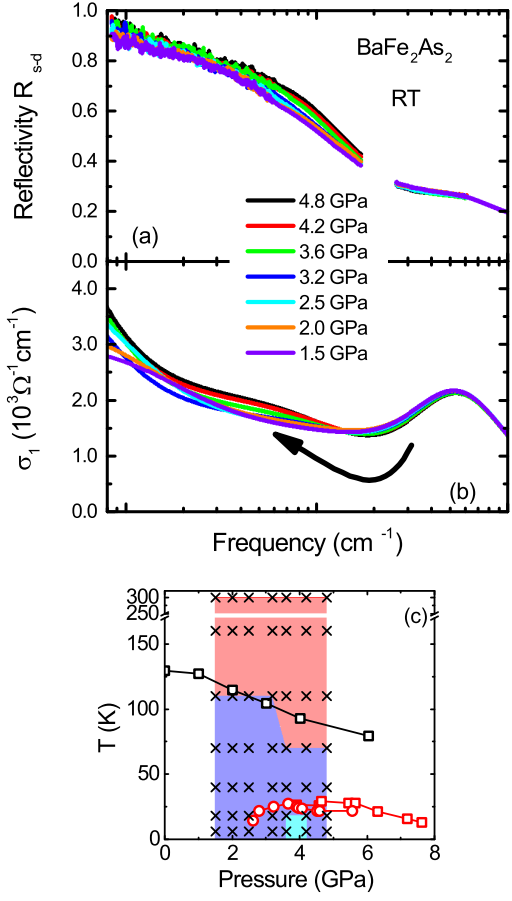}%
\caption{(color online) (a) Room-temperature (RT) refectivity spectra at various pressures. (b) RT optical conductivity spectra, obtained from Drude-Lorentz fittings of the reflectivity data, at various pressures. The arrow illustrates the pressure-induced SW transfer. (c) $P$-$T$ phase diagram for BaFe$_2$As$_2$. The open symbols are based on magnetic susceptibility and resistivity measurements \cite{Colombier2009, Alireza2009, Ahilan2008}. Crosses show the measured ($P$,$T$) points of this study. Red area: region where we did not observe any signature of SDW. Blue area: region, where a SDW state can be observed. Light blue area: region, where we observed SC signature.}%
\label{Normalstate}%
\end{figure}

The room-temperature reflectivity spectra of Ba122 are depicted in Fig.~\ref{Normalstate}(a) for various pressures. The plasma edge slightly blueshifts with increasing pressure, indicating a small increase of the charge carrier density.In Fig.~\ref{Normalstate}(b) the pressure-dependent room temperature optical conductivity spectra are shown, obtained by Drude-Lorentz simulation of the measured reflectivity data. With increasing pressure, an increase of the optical conductivity below $\sim$1500 cm$^{-1}$ and a progressive spectral weight (SW) transfer from high to low energies occurs (illustrated by the arrow in Fig.~\ref{Normalstate}(b)).
The optical conductivity curves merge in the high-energy region ($\sim$5000 cm$^{-1}$) which defines the energy range where the overall SW redistribution occurs.

\begin{figure}[t]
\centering
\includegraphics[width=1.05\columnwidth]{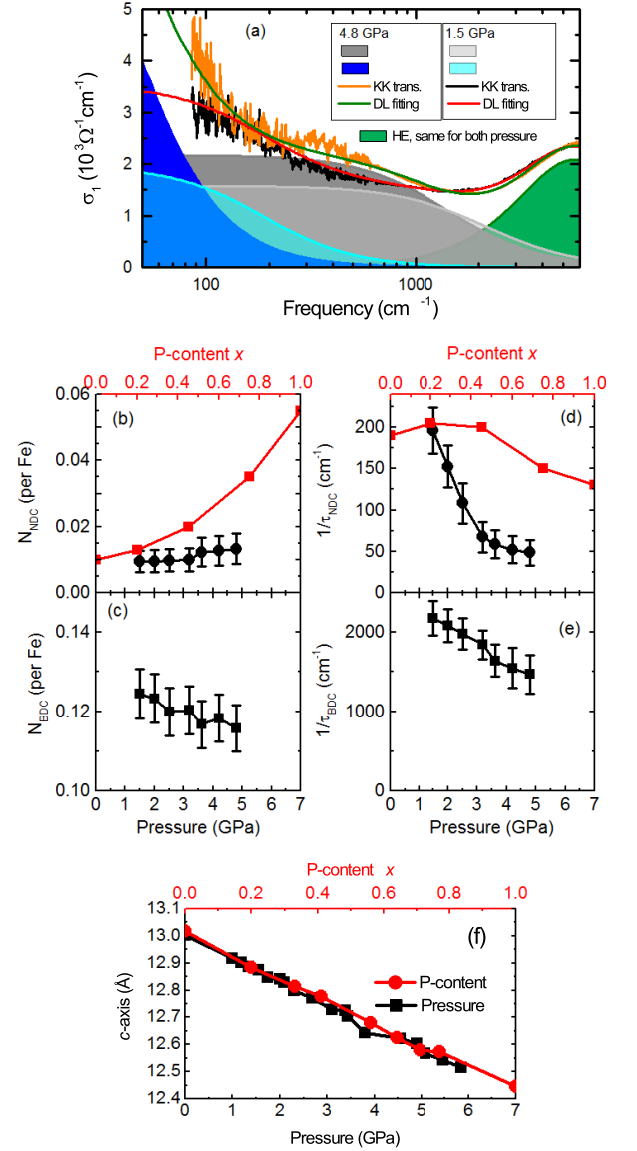}
\caption{(color online) Normal state. (a) Room-temperature optical conductivity and the components from Drude-Lorentz (DL) model fitting for 1.5 GPa and 4.8 GPa. (b, d) SW per Fe atom and scattering rate of the narrow Drude component (NDC) as a function of pressure. A comparison with P-doping (from Ref. \cite{Nakajima2013}) (red curves, x-scale is up) is also given. (c, e) SW per Fe atom and scattering rate of the broad Drude component (BDC) as a function of pressure. (f) The normalization of the external pressure \cite{Kimber2009} and P-content \citep{Kasahara2010} based on the change of the $c$-axis lattice constant.}
\label{Components}
\end{figure}

\begin{figure*}[t]%
\centering
\includegraphics[width=2\columnwidth]{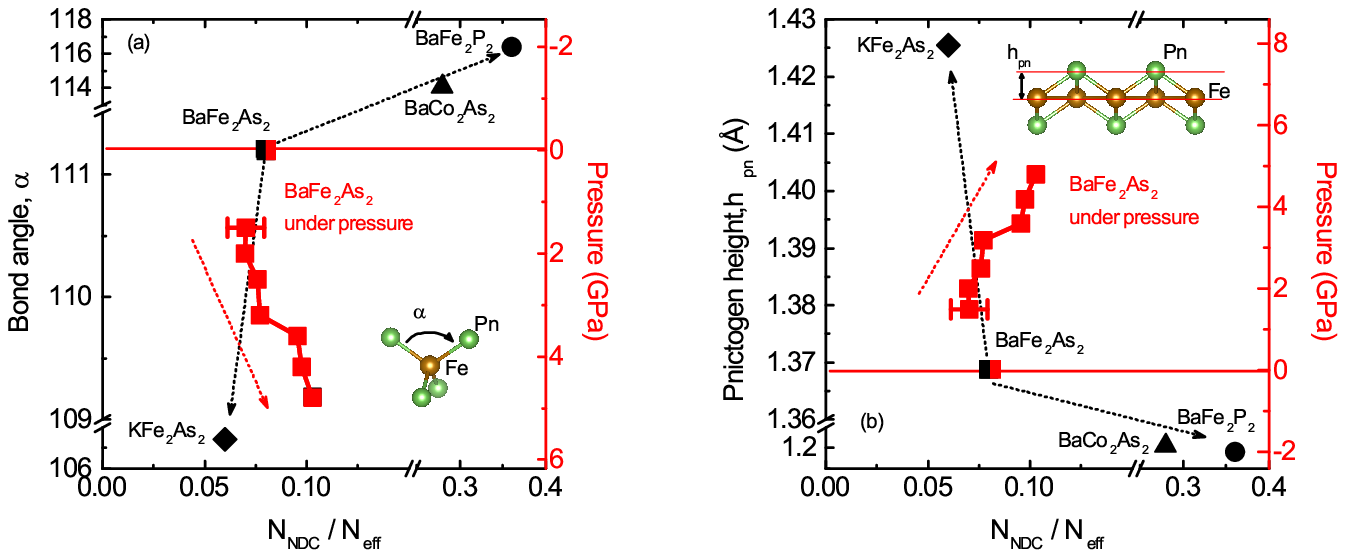}%
\caption{(color online) Structural parameters. Fraction of the narrow Drude weight, $N_{NDC}$/$N_{eff}$, as a function of bond angle, $\alpha$, (a) and pnictogen height, $h_{pn}$, (b) and pressure (a, b). The pressure data has been normalized to $\alpha$ and $h_{pn}$ at 0 and 4.8 GPa. The end materials at ambient conditions are shown as well for comparison \cite{Kimber2009, Kasahara2010, Nakajima2013}.}%
\label{Parameters}%
\end{figure*}

A detailed analysis of the optical conductivity based on Drude-Lorentz fittings is given in Figs.~\ref{Components}(a), following earlier optical studies \cite{Hu2008, Barisic2010, Nakajima2010, Nakajima2013, Nakajima2014}.
The optical conductivity spectra can be described by two Drude components in the lower energy region and one Lorentzian in the higher energy range corresponding to interband excitations. The existence of two Drude components relates to the multiband character of the iron pnictides \cite{Hu2008, Barisic2010, Nakajima2010, Nakajima2013, Nakajima2014}: A narrow, coherent Drude component with small spectral weight and a very broad Drude component (BDC) with incoherent character. With increasing pressure the narrow Drude component (NDC) is getting sharper, as illustrated by the steep decrease in the scattering rate [Fig.~\ref{Components}(d)], and slightly gains spectral weight $N_{NDC}$ [Fig.~\ref{Components}(b)]. In comparison, the SW ($N_{BDC}$) and the scattering rate of the BDC slightly decrease [Figs.~\ref{Components}(c,e)]. The pressure-induced narrowing of the BDC may indicate that some carriers gain coherence. The effective carrier number $N_{eff}$ can be estimated by the sum of the SWs of both narrow and broad Drude components, which is conserved under pressure. The weight of the NDC and the BDC components are calculated according to:

\begin{align}
N_{NDC,BDC}(\omega) = \dfrac{2m_0V}{\pi e^2}\int\limits_0^{\omega_{cutoff}}\sigma_1(\omega)d\omega \quad ,
\label{carrierdensity}
\end{align}
where $m_0$ is the free electron mass and $V$ is the unit cell volume per Fe atom. The value of $V$ as a function of pressure was taken from Ref.~\cite{Kimber2009}. Their contributions, $\sigma_1(\omega)$, were extracted from the total optical conductivity by Drude-Lorentz model fitting, and the integration according to Eq.~(\ref{carrierdensity}) was carried out with a cutoff frequency $\sim$10000 cm$^{-1}$. The covered energy range is high enough to encounter all the spectral weight of the components, and above this energy range the tail of the Drude components are negligible.

Considering the  pressure behavior of $N_{NDC}$ and $N_{BDC}$, it is reasonable to assume that the pressure-induced SW transfer from higher energies to lower energies basically occurs between these two Drude components. The narrowing of the BDC is slower than that of the NDC, indicating that pressure mainly affects the NDC rather than the BDC.

The fraction of the narrow Drude weight, $N_{NDC}$/$N_{eff}$, is a measure of the degree of coherence of the carrier dynamics and was suggested as a measure of electronic correlations \cite{Nakajima2014}. Fig.~\ref{Parameters} displays the dependence of this fraction on important structural parameters -- namely pnictogen height and Fe-As bond angle -- as well as on external pressure, including also some end materials at ambient conditions.
According to our optical data the fraction $N_{NDC}$/$N_{eff}$, and hence the coherence of the carriers, increases only slightly under pressure (Fig.~\ref{Parameters}). Accordingly, the electronic correlations remain in the strongly correlated regime for all studied pressures \cite{Nakajima2014}.

This result stands in stark contrast to the effect of isovalent P-doping, which causes a weakening of the electron correlations, i.e., the electron correlations are three times weaker in BaFe$_2$P$_2$ compared to BaFe$_2$As$_2$ (Fig.~\ref{Parameters}). P-doping has a stronger effect on the SW of the NDC than pressure [see Fig.~\ref{Components}(b)], and this might be the reason that the electronic correlations are getting weaker with increasing P-content, while it hardly changes with pressure. Interestingly, P-doping causes only a small decrease in scattering rate of the NDC, while it steeply decreases under pressure [Fig.~\ref{Components}(d)]. This indicates that the mobility of the carriers might increase faster with external pressure. Even though the mobility of the carriers increases, the electronic correlations are hardly changed, most probably due to the fact that the SW of the NDC is significantly smaller compared to the BDC. Please note that a normalization between pressure and P-content is necessary for a direct comparison. In this study we prefer to do this normalization by the decrease of the $c$-axis lattice constant with respect to ambient pressure \cite{Kimber2009} and zero P-content \cite{Kasahara2010}. This might not be the best way, however, since it has been thought that the main P-doping effect is chemical pressure, such normalization based on a lattice constant, which strongly affects the unit volume, is reasonable. It is worth to note that other normalization methods based on the phase diagram has also been tried. The discussion based on this normalization still holds and does not change greatly. We believe, that the normalization based on the $c$-axis lattice constant also convenient to compare other doping cases such as electron or hole doping, as well.

\begin{figure*}[t]%
\centering
\includegraphics[scale=1.1]{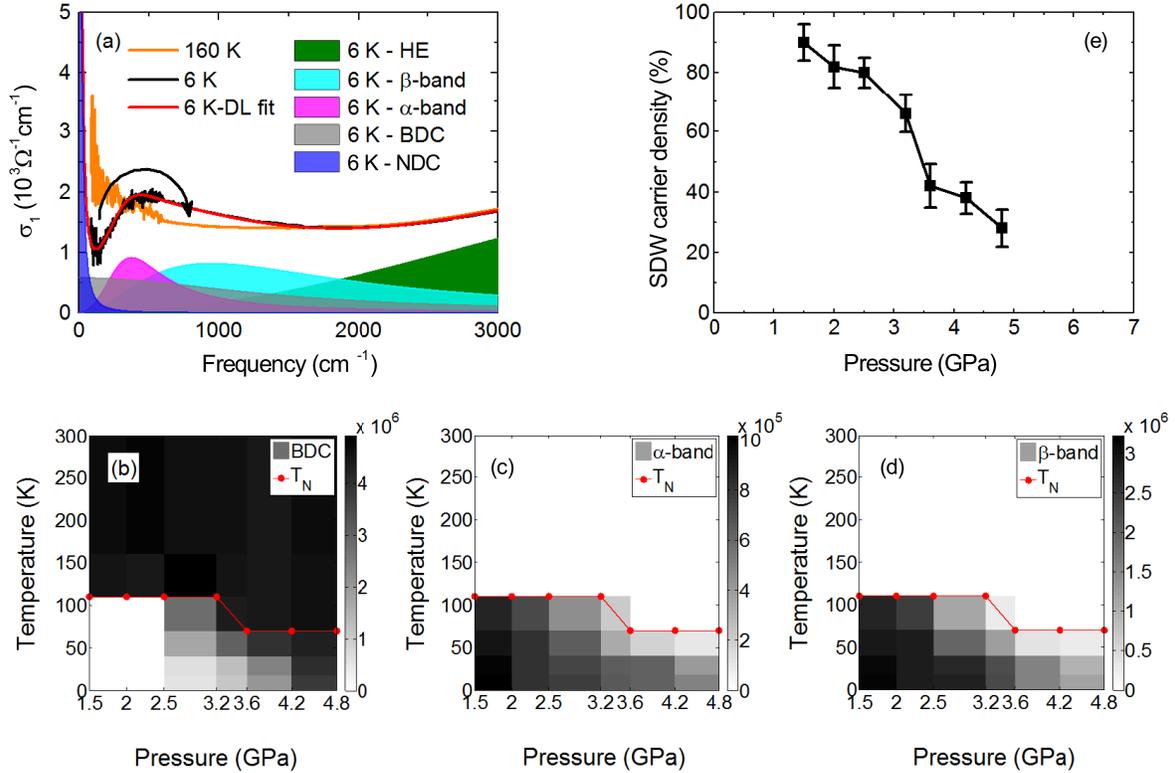}%
\caption{(color online) SDW state. (a) Optical conductivity for 160 K (above $T_N$) and 6 K at 3.2~GPa. The components from Drude-Lorentz fitting for 6 K are also given. Arrow indicates the SW transfer upon SDW gap opening. (b-d) Pressure- and temperature-dependent SW of BDC, $\alpha$- and $\beta$- bands. (e) Pressure dependence of the spin-density wave (SDW) state carrier concentration.}%
\label{SDWstate}%
\end{figure*}

Furthermore, P-doping has a ``negative'' pressure effect on the Fe-As bond angle [Fig.~\ref{Parameters}(a)] and the pnictogen height [Fig.~\ref{Parameters}(b)], like in the electron-doped case. In contrast, external pressure resembles more the hole-doped case, while still some differences do exist since the electron correlations are weakening slightly. In conclusion, the comparison demonstrates that external pressure affects the Fermi surface in these materials quite differently than isovalent doping.

With decreasing temperature, a magnetically ordered so-called SDW state is observed in BaFe$_2$As$_2$, with the suppression of the optical conductivity in the low-energy region and the transfer of the lost SW to the higher-energy region of the spectrum. Fig.~\ref{SDWstate}(a) illustrates the emergence of the SDW state at 3.2~GPa during cooling down, with the SW transfer indicated by an arrow. Below \textit{T}$_N$ the BDC is suppressed and a gap-like feature develops in the optical conductivity spectrum. Besides, a sharpening of the NDC is observed during cooling down, as expected for a system with a metallic characteristic \cite{Hu2008}.

The SDW-related gap-like feature can be described by two Lorentzian contributions, namely the $\alpha$-band at $\sim$360~cm$^{-1}$ and the $\beta$-band at $\sim$950 cm$^{-1}$ \cite{Moon2010}, which is associated with the itinerant and local magnetism, respectively. This twofold absorption feature reflects the fact that the actual magnetism in iron-pnictides has both itinerant and local character as shown by many experimental probes \cite{Johannes2009, Zhao2009, Qazilbash2009}.
In Figs.~\ref{SDWstate}(b-d) the SW of BDC, $\alpha$- and $\beta$- bands are displayed as a function of pressure and temperature. With increasing pressure both absorption bands lose spectral weight, hence the SDW state is gradually suppressed by pressure and concomitantly the BDC is recovered. According to the SW analysis of the $\alpha$- and $\beta$- bands [see Figs.~\ref{SDWstate}(c,d)] the applied pressure affects both SDW components similarly, i.e., the pressure effect on both itinerant and local magnetism is similar.

The overall plasma frequency obtained through the contributions of the NDC and BDC is an important parameter, since the plasma frequency is a measure of the charge carrier density ($\omega_p^2\propto n/m^{\ast}$, where $n$ is the carrier density and $m^{\ast}$ the effective mass). A comparison of the $\omega_p^2$ in the normal and SDW state may give a more quantitative insight into the pressure effect: $\omega_{p,40K}^2/\omega_{p, 300K}^2\approx$ 10$\%$ and 58$\%$ for 1.5 GPa and 3.6 GPa, respectively. Assuming that $m^{\ast}$ does not change significantly with temperature \citep{Hu2008}, this means that almost 90$\%$ of the free carriers contribute to the SDW state at 1.5~GPa, while this value drops to $\approx$ 40$\%$ at 3.6 GPa. The SDW carrier density as a function of pressure has been given in Fig.~\ref{SDWstate}(e). These values are also consistent with the decrease of the SW of the $\alpha$- and $\beta$- bands with increasing pressure [Fig.~\ref{SDWstate}(c,d)].

\begin{figure}[t]%
\centering
\includegraphics[scale=1]{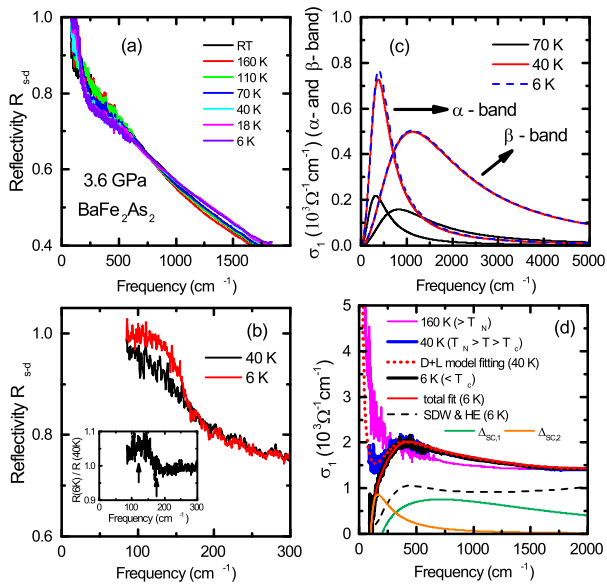}%
\caption{(color online) Main results in the SC state at 3.6 GPa. (a) Temperature-dependent reflectivity spectrum. (b) Low-energy reflectivity with the step-like SC signature at 6~K. Inset: Ratio R$_{SC}$ / R$_{normal}$, where the arrows indicate the 2$\Delta_{SC}$ for two SC gaps. (c) SDW components in the normal and SC state. (d) Optical conductivity from KK analysis at 160~K (above the SDW transition), at 40 K (just above \textit{T}$_c$) together with the Drude-Lorentz fit, and at 6 K (SC state) together with the two isotropic gaps fits (from Mattis-Bardeen approximation) and the mid-infrared components (SDW + HE Lorentzians).}
\label{SCstate}%
\end{figure}

With increasing pressure, the SDW state is gradually suppressed, and at 3.6 GPa and 4.2 GPa a SC state has emerged (Fig.~\ref{Normalstate}(c)). In this pressure range we could observe superconductivity most clearly \bibnote{Possibly the SC state does exist in other pressure ranges (below 3.6 GPa and above 4.2 GPa), as well. However, possibly, low \textit{T}$_c$/low superfluid density for these pressures push  the SC condensation energy to the lower energy region, which cannot be observed successfully in this study due to the diffraction limit related to the small samples' size.}. Fig.~\ref{SCstate} summarizes the main results for the SC state at 3.6 GPa. In Fig.~\ref{SCstate}(a) the low-energy reflectivity has been given, where one can see the small effect of the magnetic ordering starting below 70 K. When cooling below 40 K, a step-like feature that is confined to the very low-energy part of the spectrum can be seen, which is characteristic for the SC state [Fig.~\ref{SCstate}(b)]. Below 165~cm$^{-1}$ the reflectivity increases continuously towards zero frequency and reaches unity at $\sim$95 cm$^{-1}$. This clearly indicates that BaFe$_2$As$_2$ has a full-gap superconductivity tendency under pressure.

Most of the iron-pnictides exhibit a full-gap behavior in the SC region \cite{Miao2012, Terashima2009}, but some of them were shown to be nodal superconductors, among them LaFePO, P-doped BaFe$_2$As$_2$, KFe$_2$As$_2$, LiFeP, and FeSe \cite{Hashimoto2010, Hashimoto2012, Dong2010}. The points and the structure of the nodes, however, are under debate \cite{Zhang2012, Yamashita2011}.
The difference in SC gap symmetry between P-doped BaFe$_2$As$_2$ and pressurized BaFe$_2$As$_2$ (our data) is most striking, since P-doping is expected to act as a chemical pressure and therefore similar effects are expected.
Our finding suggests that the SC mechanism for isovalent P-doping and external pressure is different.

It was demonstrated experimentally that superconductivity emerges in iron pnictides when the SDW state is weakened, and the coexistence of these two states was shown by nuclear magnetic resonance \cite{Julien2009}, muon spin resonance \cite{Bernhard2012, Marsik2010}, neutron and X-ray diffraction \cite{Pratt2009}, and infrared spectroscopy (not in 122-system, but in a related system) \cite{Wang2014}. Several optical studies on the SC region have been performed (mostly focusing on the optimally doped region) \cite{Charnukha2011, Heumen2010, Wu2010, Dai2013}. However, in the so-called underdoped region, the optical studies were not sufficient to discuss the relation between the SC and SDW states: the region where the SDW and SC states coexist is not easy to observe, mainly due to the low $T_c$s and strongly weakened SDW response. In contrast, under pressure it is possible to establish such a region, where a significantly strong SDW response coexists with a SC region with high \textit{T}$_c$. The coexistence of the SC and SDW phases in BaFe$_2$As$_2$ is clearly demonstrated by the optical data of this study [see Figs.~\ref{SCstate}(c) and (d)].

The $\alpha$- and $\beta$-bands at 3.6~GPa are depicted in Fig.~\ref{SCstate}(c) for three selected temperatures, namely at 70~K (SDW state), at 40 K (SDW state) and 6 K (SC state): Between 70 K and 40 K both bands gain SW as expected, while their SW is not significantly affected by the emergence of superconductivity. From these results, two important conclusions can be drawn: Firstly, the SDW response seems to be robust against the SC transition as the SW of both bands does not decrease between 40~K and 6~K. This suggests that the SDW and the SC state are two separate orders that coexist in the SC region. Whether they are spatially separated orders or not, cannot be determined based on the current study due to the limited spatial resolution. Secondly, there is no increase of the SW of both bands either upon the SC transition, indicating that the remaining charge carriers contribute to the SC condensation rather than to the SDW state. This result might indicate a possible competition between these two states, where the SC state seems to be energetically favorable.

The optical conductivity spectra at 3.6~GPa for the normal state (160~K), SDW state (40~K), and SC state (6~K) are depicted in Fig.~\ref{SCstate}(d) together with the corresponding fittings. The SDW response with the low- to high-energy SW transfer can be seen at 40~K. At 6~K, below $\sim$ 200 cm$^{-1}$ a gradual suppression of the optical conductivity as the indication of the SC condensation occurs, while the high-energy SW remains the same indicating that the SDW state does not disappear upon the SC transition. The optical conductivity in the SC region was fitted with the Mattis-Bardeen approximation with two-isotropic gaps \cite{Mattis1958, Zimmermann1991} and the estimated SC gaps are $\Delta_{SC,1} \approx 7.5$ meV and $\Delta_{SC,2} \approx 10.9$ meV. We assumed the two gap signature due to the fact that we have two Drude-components in the normal state. Hence a superconducting gap opening occur in each component. These values are reasonably similar to the other systems with similar \textit{T}$_c$ \cite{Inosov2011}.

According to Ferrell-Glover-Tinkham (FGT) sum rule, the difference of the spectral weight between \textit{T} $\approx$ \textit{T}$_c$ and \textit{T} $<<$ \textit{T}$_c$ is so-called missing area is related with the superconducting carrier density as:

\begin{align}
\omega^2_{p,SC} = \frac{120}{\pi}\int\limits_0^{\omega_{cutoff}} [\sigma_1(\omega, \textit{T} \approx \textit{T}_c) - \sigma_1(\omega, \textit{T} << \textit{T}_c)]d\omega
\label{SCD}
\end{align}

$\omega_{cutoff}$ is chosen as $\sim$ 5000 cm$^{-1}$. The SC plasma frequency at 6 K (\textit{T} = 40~K $\approx$ \textit{T}$_c$) for this pressure range that is calculated with the missing area method amounts to $\omega_{p,SC}\approx$ 4650~cm$^{-1}$, where the calculated penetration depth is $\approx$ 3000~\r{A} $\pm$ 300 \r{A}. Please note that this penetration depth value is very close to other 122-family iron pnictides \citep{Wu2010a, Wu2011}. Another way to calculate the superconducting carrier density is through $\sigma_2(\omega) = \omega^2_{p,SC}/4\pi\omega$. The $\omega_{p,SC}$ calculated with this method is $\approx$ 5120 cm$^{-1}$. The two different method give fairly similar values.

The value of $\omega_p$ at 40 K is calculated to be $\approx$ 9000 cm$^{-1}$ from the fitting parameters of the two Drude components ($\omega^2_{p} \propto  n/m^\ast$). If we compare the superconducting carrier density, $\omega^2_{p,SC} / \omega^2_{p,40K}$, we find that $\approx$ 27$\%$ of the carriers contribute to the superconducting condensation. Moreover, a very clear superconducting gap signature has been observed for this sample, which is not expected in the case of the clean limit superconductivity \citep{Lobo2015}. Therefore, at first glance, the Mattis-Bardeen approximation used to determined the SC energy gap is reasonable, since the system shows a dirty-limit superconductivity \bibnote{We note here, that in principle such kind of approximation for the SC state needs to take into account the coexisting SDW state in the case of the pnictides materials.}.

Finally, we would like to make a comment on the filamentary superconductivity induced by uniaxial stress in iron pnictide superconductors. Owing to the fact that a quasi-hydrostatic pressure transmitting medium has been used in this study, the question arises, whether the superconductivity is induced by uniaxial stress or not. As the ruby luminescence spectra given in Fig.~\ref{hydrostaticity} demonstrate, the ruby R1 line is sharp and the difference between two locations is small. Moreover, it has been discussed previously that the magnetically ordered state tends to be suppressed much more quickly in the case of an unhydrostatic environment \citep{Yamazaki2010}. In our study, we can observe the magnetically ordered state even at the highest applied pressure. As a result, even though we cannot completely rule out the possibility, it is highly unlikely that the observed superconductivity is filamentary under these conditions.

\section{Conclusions}

In summary, we performed temperature- and pressure-dependent reflectivity measurements on BaFe$_2$As$_2$ single crystals. With increasing pressure, the suppression of the SDW state and the emerging of the SC state was observed at 3.6 and 4.2 GPa. At these pressures, the SC and SDW states coexist. The pressure-induced SC region shows a full gap tendency, in contrast to the isovalent P-doping case.  Together with the observed opposite effect of P-doping and external pressure on  the pnictogen height and the Fe-pnictogen bond angle, this suggests that the driving mechanism of superconductivity may not be the same in these two cases. We also demonstrated that pressure does not have a strong effect on the electron correlations in the system in contrast to the P-doping case. In fact, our results show a behavior of BaFe$_2$As$_2$ under pressure very similar to the one for hole-doping in Ba$_{1-x}$K$_x$Fe$_2$As$_2$ \cite{Kimber2009, Nakajima2014}: Besides important structural parameters the electron correlations are affected in a similar fashion. Our findings suggest that external pressure mostly affects the hole bands of the Fermi surface.

\section{Acknowledgements}

We thank Masamichi Nakajima for the UV-region reflectivity data. We also would like to acknowledge fruitful discussions with Roser Valent\'{i}, Harald Jeschke, and Milan Tomi\'{c}. This work is financially supported by the Bavarian Research Foundation and the BMBF. The work in Japan is supported by IRON-SEA, JST.

$^*$ ece.uykur@physik.uni-augsburg.de

$^\dagger$ christine.kuntscher@physik.uni-augsburg.de

\bibliographystyle{apsrev4-1}
\bibliography{References}

\end{document}